\documentclass{svmult}

\title*{
Optimal investment with bounded VaR for
power utility functions\thanks {This work was supported by the
scinetific cooperation CNRS/DPGRF, Projet DZAC 19856, France-Alg\'erie.
The second author is partially supported by the RFBR-Grant 09-01-00172-a.
} }
\titlerunning{Optimal investment with bounded VAR}
\author{B\'enamar Chouaf and Serguei Pergamenchtchikov}
\institute{ B\'enamar Chouaf  \at 
Universit\'e de Sidi Bel Abbes
Laboratoire de Math\'ematiques Appliqu\'ees
\at
\email{bchouaf@univ-sba.dz}
 \and Serguei Pergamenchtchikov \at
Laboratoire de Math\'ematiques Rapha\"el Salem,
 Universit\'e de Rouen,   BP.12,
76801 Saint Etienne du Rouvray, France,
\email{Serge.Pergamenchtchikov@univ-rouen.fr}  }

\usepackage{mathptmx}       
\usepackage{helvet}         
\usepackage{courier}        
\usepackage{srcltx}
\usepackage{amssymb}
\usepackage{amsfonts}
\usepackage{amsmath}

\usepackage{graphicx}        
\usepackage{multicol}        

\newcommand{\beao}{\begin{eqnarray*}}
\newcommand{\eeao}{\end{eqnarray*}\noindent}

\newcommand{\beam}{\begin{eqnarray}}
\newcommand{\eeam}{\end{eqnarray}\noindent}

\def\bbr{{\mathbb R}}

\newcommand{\sign}{\mbox{sign}}

\newcommand{\si}{{\sigma}}

\newcommand{\ov}{\overline}
\newcommand{\wh}{\widehat}
\newcommand{\wt}{\widetilde}

\newcommand\cE{{\cal E}}

\newcommand\cF{{\cal F}}

\newcommand\cN{{\cal N}}

\newcommand\cY{{\cal Y}}

\def\text#1{\hbox{#1}}

\newcommand{\halmos}{\quad\hfill\mbox{$\Box$}}

\def\E{{\bf E}}
\def\P{{\bf P}}
\def\D{{\bf D}}
\def\L{{\bf L}}

\def\Chi{{\bf 1}}
\def\d{{\rm d}}
\def\build #1_#2{\mathrel{\mathop{\kern 0pt #1}\limits_\zs{#2}}}
\newcommand{\zs}[1]{{\mathchoice{#1}{#1}{\lower.25ex\hbox{$\scriptstyle#1$}}
{\lower0.25ex\hbox{$\scriptscriptstyle#1$}}}}

\numberwithin{equation}{section}

\begin{document}

\maketitle

\abstract{We consider the optimal investment problem
for Black-Scholes type financial market with bounded VaR
measure on the whole investment interval $[0,T]$. The explicit form for
the optimal strategies is found.
}


\vspace*{2mm}


\section{Introduction}\label{sec:1}

We consider an investment problem 
aiming at  optimal
terminal wealth at maturity $T$.  The classical approach to this
problem goes back to Merton~\cite{Me}
 and involves utility functions,
more precisely, the expected utility serves as the functional which
has to be optimized.

We adapt this classical utility maximization approach to nowadays
 industry practice:
investment firms customarily impose limits on the risk of trading
 portfolios.
These limits are specified in terms of downside Value-at-Risk (VaR) risk measures.

As Jorion~\cite{Jo}, p. 379  points out, VaR creates a common
 denominator for the comparison of different risk activities. Traditionally,
 position limits of traders are set in terms of notional exposure, which may
 not be directly comparable across treasuries with different
 maturities.
In contrast, VaR provides a common denominator to compare various
asset
 classes and business units.
The popularity of VaR as a risk measure has been endorsed by
 regulators, in particular, the Basel Committee on Banking Supervision, which
 resulted in mandatory regulations worldwide.

Our approach combines the classical utility maximization with risk
 limits in terms of VaR.
This leads to control problems under restrictions on uniform
versions
 of VaR, where the risk bound is supposed to be intact throughout the
 duration of the investment.
To our knowledge such problems have only been considered in dynamic
 settings, which reduce intrinsically to static problems.
Emmer, Kl\"uppelberg and Korn~\cite{EmKlKo} consider a dynamic
market, but maximize only the
 expected wealth at maturity
under a downside risk bound at maturity. Basak and Shapiro
\cite{BaSha} solve the utility optimization problem
 for complete markets with bounded VaR at maturity.
Gabih, Gretsch and Wunderlich \cite{GaGrWu} solve the utility
optimization problem for constant coefficients markets with bounded
ES at maturity. Kl\"uppelberg and Pergamenshchikov 
\cite{KlPe1}-\cite{KlPe2} considered the optimisation problems
with bounded Var and ES risk measure on the whole time interval
in the class of the nonrandom financial stratedies. 
In this paper we consider the optimal investment problem with
the 
bounded VaR uniformly on whole time interval $[0,T]$ for 
all admissible financial strategies (nonrandom or random).
It should be noted that it is immpossible to calculate the explicit
form of the VaR risk measure for the random financial strategies. This is the main
difficulty in such problems.
In this paper we propose some method to overcome this difficulty
by applying  optimisations methods in the Hilbert spaces. We find 
the explicit form for the optimal strategies. 

Our paper is organised as follows. In Section~\ref{sec:Mo} we
formulate  the Black-Scholes
model for the price
 processes. 
In Section~\ref{sec:Op} all optimization
problems and their solutions
 are given.
 All proofs
are summarized in Section~\ref{sec:Pr}
 with the technical lemma postponed to
the Appendix~\ref{sec:A}.

\section{The model}\label{sec:Mo}

We consider a Black-Scholes type financial market consisting of one
{\em riskless bond} and several {\em risky stocks}. Their respective
prices $(S_\zs{0}(t))_\zs{t\ge 0}$ and $(S_\zs{i}(t))_\zs{t\ge 0}$
for $i=1,\ldots,d$ evolve according to the equations:
\begin{equation}\label{sec:Mo.1}
\left\{\begin{array}{ll} \d
S_\zs{0}(t)\,=\,r_\zs{t}\,S_\zs{0}(t)\,\d t\,, &
 S_\zs{0}(0)\,=\,1\,,\\[5mm]
\d S_\zs{i}(t)\,=\,S_\zs{i}(t)\,\mu_\zs{i}(t)\,\d
t\,+\,S_\zs{i}(t)\, \sum^d_\zs{j=1}\,\sigma_\zs{ij}(t)\,\d
W_\zs{j}(t)\,, & S_\zs{i}(0)\,=\,s_i\,>0\,,
\end{array}\right.
\end{equation}
Here $W_\zs{t}=(W_\zs{1}(t),\ldots,W_\zs{d}(t))'$ is a standard
$d$-dimensional Brownian motion; $r_\zs{t}\in\bbr$ is the {\em
riskless interest rate},
$\mu_\zs{t}=(\mu_\zs{1}(t),\ldots,\mu_\zs{d}(t))'\in\bbr^d$ is the
 vector of
{\em stock-appreciation rates} and
$\sigma_\zs{t}=(\sigma_\zs{ij}(t))_\zs{1\le i,j\le d}$ is the matrix
of {\em stock-volatilities}. We assume that the coefficients
$r_\zs{t}$, $\mu_\zs{t}$ and
 $\sigma_\zs{t}$
are deterministic functions, which are right continuous with  left
limits (c\`adl\`ag). We also assume that the matrix $\sigma_\zs{t}$
is non-singular for Lebesgue-almost all $t\ge0$.

We denote  by $\cF_\zs{t}=\sigma\{W_\zs{s}\,,s\le t \}$, $t\ge 0$,
the
 filtration generated by the Brownian motion (augmented by the null
 sets).
Furthermore, $|\cdot|$ denotes the Euclidean norm for vectors and
the
corresponding matrix norm for matrices. 

For $t\ge 0$ let $\phi_\zs{t}\in\bbr$ denote the amount of investment
into
 bond and
$$
\varphi_\zs{t}=(\varphi_\zs{1}(t),\ldots,\varphi_\zs{d}(t))'\in\bbr^d
$$
the amount of investment into risky assets. We recall  that a {\em
trading strategy} is an $\bbr^{d+1}$-valued $(\cF_\zs{t})_\zs{t\ge
 0}$-progressively measurable process
$(\phi_\zs{t},\varphi_\zs{t})_\zs{t\ge 0}$ and that
$$
X_\zs{t}\,=\,\phi_\zs{t}\,S_\zs{0}(t)\,+
\sum^d_\zs{j=1}\,\varphi_\zs{j}(t)\,S_\zs{j}(t)\,,\quad t\ge 0\,,
$$
is called the {\em wealth process}.

The trading strategy $((\phi_\zs{t},\varphi_\zs{t}))_\zs{t\ge 0}$
is called {\em self-financing}, if the
wealth
 process
 satisfies the following equation
\begin{equation}\label{sec:Mo.2}
X_\zs{t}\,=\,x\,+\, \int^t_0\,\phi_\zs{u}\,\d S_\zs{0}(u)\,+
\,\sum^d_\zs{j=1}\,\int^t_0\,\varphi_\zs{j}(u)\,\d S_\zs{j}(u)
\,, \quad t\ge 0\,,
\end{equation}
where $x>0$ is the initial endowment.

In this paper we work with relative quantities,  i.e., we define for $
 j=1,\ldots,d$
$$
\pi_\zs{j}(t)\,:=\,
\frac{\varphi_\zs{j}(t)\,S_\zs{j}(t)}
{\phi_\zs{t}\,S_\zs{0}(t)+\sum^d_\zs{j=1}\,\varphi_\zs{i}(t)\,S_\zs{i}(t)}\,,\quad t\ge 0\,.
$$
Then $\pi_\zs{t}=(\pi_\zs{1}(t),\ldots, \pi_\zs{d}(t))'$, $t\ge 0$,
is
 called
the {\em portfolio process} and we assume throughout that it is
$(\cF_\zs{t})_\zs{t\ge 0}$-progressively measurable. We assume that
for the fixed investment horizon $T>0$
$$
\|\pi\|_T^2:= \int_0^T |\pi_\zs{t}|^2 \d t <\infty\quad {\rm a.s.}\,.
$$
We also define with $\Chi=(1,\ldots,1)'\in\bbr^d$ the quantities
\begin{equation}\label{sec:Mo.3}
y_\zs{t}=\sigma'_\zs{t}\pi_\zs{t}\quad\mbox{and}\quad
\theta_\zs{t}=\sigma^{-1}_\zs{t}(\mu_\zs{t}-r_\zs{t}\,\Chi)\,,\quad
 t\ge 0\,,
\end{equation}
where it suffices that these quantities are defined for
Lebesgue-almost
 all $t\ge 0$.
Taking these definitions into account we rewrite equation
\eqref{sec:Mo.2}
 for $X_\zs{t}$ as
\begin{equation}\label{sec:Mo.4}
\d X_\zs{t}\,=\,X_\zs{t}\,(r_\zs{t}\,+\,y'_\zs{t}\,\theta_\zs{t})
\,\d t\,+\,X_\zs{t}\,y'_\zs{t}\,\d
W_\zs{t}\,,\quad
 X_\zs{0}\,=\,x>0\,.
\end{equation}
This implies in particular that any optimal investment strategy is
 equal to
$$\pi^*_\zs{t}=\sigma_\zs{t}'^{-1}y^*_\zs{t}\,,
$$
 where $y^*_\zs{t}$ is the optimal control
 process
for equation \eqref{sec:Mo.4}. We also require for the investment horizon
$T>0$
\begin{equation}\label{sec:Mo.5}
\|\theta\|^2_\zs{T}=\int^T_\zs{0}\,|\theta_\zs{t}|^2
\,\d t\,<\,\infty\,.
\end{equation}
We assume that $(y_\zs{t})_\zs{0\le t\le T}$ is any 
$(\cF_\zs{t})_\zs{0\le t\le T}$ - adapted
a.s. square integrated process, i.e.
$$
\|y\|^2_\zs{T}=
\int^T_\zs{0}|y_\zs{t}|^2\,\d t\,<\,\infty
\quad\mbox{a.s.,}
$$
such that the stochastic equation \eqref{sec:Mo.4}
has a unique strong solution. We denote by $\cY$
the class of all such processes $y=(y_\zs{t})_\zs{0\le t\le T}$. Note that for every $y\in\cY$, through It\^o's formula, we represent the equation
\eqref{sec:Mo.4} in the following form
(to emphasize that the wealth
process corresponds to some control process $y$ we write
$X^{y}$)

\begin{equation}\label{sec:Mo.6}
 X^{y}_\zs{t}\,=\,x\,e^{R_\zs{t}+(y,\theta)_\zs{t}}\,\cE_\zs{t}(y)\,,
\end{equation}
where $R_\zs{t}=\int^t_0 r_\zs{u}\d u$, 
$(y,\theta)_\zs{t}=\int^t_0 y'_\zs{u}\,\theta_\zs{u}\d u$
and the process $(\cE_\zs{t}(y))_\zs{0\le t\le T}$ is the stochastic exponent for $y$, i.e.
$$
\cE_\zs{t}(y)= \exp\Big(\int^t_0 y'_\zs{u}\d W_\zs{u} -\frac{1}{2}
\int^t_0 |y_u|^2\d u\Big)\,.
$$
Therefore, for every $y\in\cY$ the process
$(X^{y}_\zs{t})_\zs{t\ge0}$ is a.s. positive and continuous.

For initial endowment $x>0$ and a control process
$y=(y_\zs{t})_\zs{t\ge 0}$
 in $\cY$,
we introduce the {\em cost function}
\begin{equation}\label{sec:Mo.7}
J(x,y):=\E_\zs{x}\,\left(X^{y}_\zs{T}\right)^{\gamma}\,,
\end{equation}
where $\E_\zs{x}$ is the expectation operator conditional on
  $X^{y}_\zs{0}=x$. 

For $0<\gamma<1$ the utility function $U(z)=z^\gamma$ is concave and
is
 called a power (or HARA) utility function.
We include the case of $\gamma=1$, which corresponds to simply
 optimizing expected consumption and terminal wealth. In combination with a
 downside risk bound this allows us in principle to disperse with the
 utility function, where in practise one has to choose the parameter $\gamma$.

\section{Optimisation problems}\label{sec:Op}
\subsection{The Unconstrained Problem}\label{subsec:Op-1}
  
We consider two regimes with cost functions \eqref{sec:Mo.7} for   
$0<\gamma<1$  and for $\gamma=1$.

\bigskip  

\begin{equation}\label{sec:Op.1}  
\max_\zs{y\in\cY}\,J(x,y)\,.  
\end{equation}  

\noindent 
First we consider  Problem~\ref{sec:Op.1} for $0<\gamma<1$.
The following result can be found in Example~6.7 on page 106 in Karatzas and Shreve 
\cite{KaSh2}; it's proof there is based by the martingale method. 

\begin{theorem}\label{Th.sec:Op.1}  
Consider Problem~\ref{sec:Op.1} for $0<\gamma<1$. 
The optimal value of $J(x,y)$  is given by   
$$
J^*(x)\,=\,\max_\zs{y\in\cY}\,J(x,y)\,=\,J(x,y^*)\,  
=\,x^{\gamma}\,\exp\{\gamma R_\zs{T}+\frac{\gamma}{2(1-\gamma)}
\|\theta\|^2_\zs{T}\} \,,
$$  
where the optimal control $y^*=(y^*_\zs{t})_\zs{0\le t\le T}$ 
is for all $0\le t\le T$ of the form  
\begin{equation}\label{sec:Op.3}
y^*_\zs{t}=\dfrac{\theta_\zs{t}}{1-\gamma} \quad  
\left(  
\pi^*_\zs{t}=\dfrac{(\si_t\si_t')^{-1}(\mu_t-r_t{\bf 1} )}{1-\gamma}  
\right)\,.
\end{equation}  
The optimal wealth process $(X^*_t)_{0\le t\le T}$ is given by  
\begin{equation}\label{sec:Op.4}
\d X^*_\zs{t}\,=\,X^*_\zs{t}\left(r_\zs{t}+\frac{|\theta_\zs{t}|^2}{1-\gamma}\right)\d t  
+X^*_\zs{t}\frac{\theta'_\zs{t}}{1-\gamma}\,\d W_\zs{t}\,, \quad   
  X^*_\zs{0}=x\,.  
\end{equation}  
\end{theorem}  
\medskip 
Let now $\gamma=1$.

\begin{theorem}\label{Th.sec:Op.2}  \cite{KlPe1}
Consider the problem~\ref{sec:Op.1} with $\gamma=1$.  
Assume a riskless interest rate  
$r_\zs{t}\ge 0$ for all $t\in [0,T]$.
If $\|\theta\|_\zs{T}>0$ then  
$$  
\max_\zs{y\in\cY}\,J(x,y)\,=\,\infty\,.  
$$  
If $\|\theta\|_\zs{T}=0$ then a solution exists and the optimal value of   
 $J(x,y)$ is  
given by  
$$
\max_\zs{y\in\cY}\,J(x,y)\,=\,J(x,y^*)\,=\,x\,e^{R_\zs{T}}\,,  
$$
corresponding  to arbitrary deterministic   
square integrable function $(y^*_\zs{t})_\zs{0\le t\le T}$. 
In this case   the optimal wealth process  
$(X^*_t)_{0\le t\le T}$ satisfies the following equation  
\begin{equation}\label{sec:Op.2}
\d X^*_\zs{t}=X^*_\zs{t} r_\zs{t}\d t\,+\,X^*_\zs{t}\,y^*_{\zs{t}}\,'\,  
\d W_\zs{t}\,,\quad X^*_\zs{0}\,=\,x\,.  
\end{equation}  
\end{theorem}  
\subsection{The Constrained Problem}\label{subsec:Op-2}

As risk measures we use modifications of the Value-at-Risk 
 as introduced in Emmer, Kl\"uppelberg and
 Korn~\cite{EmKlKo}. They can be summarized under the notion of Capital-at-Risk as
 they reflect the required capital reserve.
To avoid non-relevant cases we consider only $0<\alpha<1/2$. We use here the definition
as in \cite{KlPe1}-\cite{KlPe2}.

\begin{definition}\label{De.3.1} {\em [Value-at-Risk (VaR)]}\\
Define for initial endowment $x>0$, a control process $y\in\cY$
and
 $0<\alpha\le 1/2$ the {\em Value-at-Risk (VaR)} by
$$
{\rm VaR}_\zs{t}(x,y,\alpha):=x\,e^{R_\zs{t}}-Q_\zs{t}\,, \quad t\ge 0\,,
$$
where $Q_\zs{t}=Q_\zs{t}(x,y,\alpha)$ is the 
$(\cF^{y}_\zs{t})=\sigma\{y_\zs{s}\,,0\le s\le t\}$ measurable
random variable such that
\begin{equation}\label{sec:Op.5}
\alpha\quad\mbox{quantile of the ratio}\quad
\wh{X}^{y}_\zs{t}
=
\frac{X^{y}_\zs{t}}{Q_\zs{t}}
\quad\mbox{is equal to}
\quad 1
\end{equation}
i.e.
$$
\inf\{z\ge 0\,:\,
\P(\wh{X}^{y}_\zs{t}\,\le\,z)\ge \alpha\}
=1\,.
$$
\end{definition}

\begin{remark}\label{Re.sec:3.1}
Note that for the nonrandom financial strategies $(y_\zs{t})_\zs{0\le t\le T}$
the process $Q_\zs{t}$ is the usual $\alpha$- quantile for
the process $X^{y}_\zs{t}$. To define the ``random`` quantile for 
the process $X^{y}_\zs{t}$ we consider the ratio 
process $\wh{X}^{y}_\zs{t}$ for which the $\alpha$-
quantile is equal to $1$.
\end{remark}

\begin{corollary}\label{Co.3.1}
For every $y\in\cY$
with $\|y\|_\zs{t}>0$ the process 
$Q_\zs{t}$ defined in  Definition~\ref{De.3.1},  is given by
$$
Q_\zs{t}=x
\exp\left(R_\zs{t}+(y,\theta)_\zs{t}-\frac{1}{2}\|y\|^2_\zs{t}
+\tau_\zs{t}\|y\|_\zs{t}\right)\,, \quad t\ge 0\,,
$$
where $\tau_\zs{t}=\tau_\zs{t}(\alpha,y)$ is the $\alpha$-quantile of 
the normalized stochastic integral 
$$
\xi_\zs{t}(y)=\frac{1}{\|y\|_\zs{t}}\int^t_\zs{0}y'_\zs{u}\d W_\zs{u}\,,
$$
i.e.
\begin{equation}\label{sec:Op.6}
\tau_\zs{t}=
\inf\{z\ge -\infty :
\P\left(\xi_\zs{t}(y)\,\le\,z\right)\ge \alpha\}\,.
\end{equation}
\end{corollary}
\noindent It is clear that for any nonrandom function $(y_\zs{t})_\zs{0\le t\le T}$
the random variable 
$$
\xi_\zs{t}\sim\cN(0,1)\,,
$$
i.e. in this case 
$\tau_\zs{t}=-|z_\zs{\alpha}|$, where $z_\alpha$ is the $\alpha$-quantile of the standard normal
 distribution. 

Indeed, to obtain the explicit form for the optimal solutions
in this paper we work with a upper bound for VaR risk measure, i.e. we  consider
the 
\begin{equation}\label{sec:Op.7}
{\rm VaR}^*_\zs{t}(x,y,\alpha):=x\,e^{R_\zs{t}}-Q^*_\zs{t}\,, \quad t\ge 0\,,
\end{equation}
where
$$
Q^*_\zs{t}=x
\exp\left(R_\zs{t}+(y,\theta)_\zs{t}-\frac{1}{2}\|y\|^2_\zs{t}
+\tau^*_\zs{t}\|y\|_\zs{t}\right)
\quad\mbox{with}\quad
\tau^*_\zs{t}=\min(z_\zs{\alpha},\tau_\zs{t})\,.
$$
Obviously, 
$$
{\rm VaR}_\zs{t}(x,y,\alpha)
\le
{\rm VaR}^*_\zs{t}(x,y,\alpha)\,.
$$

\bigskip

\noindent We define the {\em level risk function} for some coefficient
 $0<\zeta<1$ as
\begin{equation}\label{sec:Op.8}
\zeta_\zs{t}(x)\,=\,\zeta\,x\,e^{R_\zs{t}}\,,\quad t\in [0,T]\,.
\end{equation}
We consider only controls $y\in\cY$ for which the Value-at-Risk is
a.s. bounded by this level function  over the interval
$[0,T]$; i.e. we require
\begin{equation}\label{sec:Op.9}
\sup_\zs{0\le t\le T}
\frac{{\rm  VaR}^*_\zs{t}(x,y,\alpha)}{\zeta_\zs{t}(x)}\,\le\,1
\quad
\mbox{a.s..}
\end{equation}

\noindent The optimisation problem is 

\begin{equation}\label{sec:Op.10}
\max_\zs{y\in\cY}\,J(x,y)\quad   
\mbox{subject to}\quad\sup_\zs{0\le t\le T}\,  
\frac{{\rm VaR}^*_t(x,y,\alpha)}{\zeta_t(x)}\,\le\,1
\quad\mbox{a.s..}
\end{equation}
To describe the optimal strategies we need the following 
function
\begin{equation}\label{sec:Op.11}
g(a):=\sqrt{2a+\wt{z}^2_\zs{\alpha}}
-\wt{z}_\zs{\alpha}
\end{equation}
with 
$$
\wt{z}_\zs{\alpha}= |z_\zs{\alpha}|-\|\theta\|_\zs{T}
\quad\mbox{and}\quad
0\le a\le a_\zs{max}:=-\ln(1-\zeta)\,.
$$
Moreover, we set
\begin{equation}\label{sec:Op.12}
a_\zs{0}=\frac{\|\theta\|^{2}_\zs{T}}{2(1-\gamma)^{2}}
+
\wt{z}_\zs{\alpha}\,
\frac{\|\theta\|_\zs{T}}{1-\gamma}
\,.
\end{equation}

\medskip

\begin{theorem}\label{Th.sec:Op.3}  
Consider the problem \eqref{sec:Op.10} for $0<\gamma<1$. Assume that 
$|z_\zs{\alpha}|\ge 2\|\theta\|_\zs{T}$.
Then 
the optimal value for the cost function
 is given by  
\begin{equation}\label{sec:Op.13}  
J(x,y^*)=x^\gamma\,
e^{\gamma R_\zs{T}+\gamma G(g^{*})}\,,
\end{equation}  
where $G(g)=g\|\theta\|_\zs{T}+(1-\gamma)g^{2}/2$,
$g^*=g(a^*)$ with
\begin{equation}\label{sec:Op.14}
a^*=\min(a_\zs{0},a_\zs{max})\,,
\end{equation} 
and the optimal control $y^*$ is for all $0\le t\le T$ of the form  
\begin{equation}\label{sec:Op.15}  
y^*_\zs{t}=
\frac{g^{*}}{\|\theta\|_\zs{T}}
\theta_\zs{t}\,\Chi_\zs{\{\|\theta\|_\zs{T}>0\}}\,.
\end{equation}
Moreover, if 
$\|\theta\|_\zs{T}>0$
then 
the optimal wealth process $(X^*_t)_{0\le t\le T}$ 
is given by  
\begin{equation}\label{sec:Op.16}
\d X^*_\zs{t}\,=\,X^*_\zs{t}\left(r_\zs{t}+
\frac{g^{*}|\theta_\zs{t}|^2}{\|\theta\|_\zs{T}}
\right)\d t  
+X^*_\zs{t}\frac{g^{*}}{\|\theta\|_\zs{T}}\,\theta'_\zs{t}\d W_\zs{t}
\quad\mbox{with}\quad
  X^*_\zs{0}=x\,;
\end{equation}  
and if $\|\theta\|_\zs{T}=0$ then $X^*_\zs{t}\,=\,x\,e^{R_\zs{t}}$ for $0\le t\le T$.
\end{theorem}

\bigskip

\begin{theorem}\label{Th.sec:Op.4}  
Consider the problem \eqref{sec:Op.10} for $\gamma=1$. Assume that 
$|z_\zs{\alpha}|\ge 2\|\theta\|_\zs{T}$.
Then 
the optimal value for the cost function
 is given by  
\begin{equation}\label{sec:Op.17}  
J(x,y^*)=x\,
e^{R_\zs{T}+ g(a_\zs{max})\|\theta\|_\zs{T}}\,,
\end{equation}  
where
and the optimal control $y^*$ is for all $0\le t\le T$ of the form  
\begin{equation}\label{sec:Op.18}  
y^*_\zs{t}=
\frac{g(a_\zs{max})}{\|\theta\|_\zs{T}}
\theta_\zs{t}\,\Chi_\zs{\{\|\theta\|_\zs{T}>0\}}\,.
\end{equation}
Moreover, if 
$\|\theta\|_\zs{T}>0$
then 
the optimal wealth process $(X^*_t)_{0\le t\le T}$ 
is given by  
\begin{equation}\label{sec:Op.19}
\d X^*_\zs{t}\,=\,X^*_\zs{t}\left(r_\zs{t}+
\frac{g(a_\zs{max})|\theta_\zs{t}|^2}{\|\theta\|_\zs{T}}
\right)\d t  
+X^*_\zs{t}\frac{g(a_\zs{max})}{\|\theta\|_\zs{T}}\,\theta'_\zs{t}\d W_\zs{t}
\quad\mbox{with}\quad
  X^*_\zs{0}=x\,;
\end{equation}  
and if $\|\theta\|_\zs{T}=0$ then $X^*_\zs{t}\,=\,x\,e^{R_\zs{t}}$ for $0\le t\le T$.
\end{theorem} 

\bigskip

\section{Proofs}\label{sec:Pr}

\subsection{Proof of Theorem~\ref{Th.sec:Op.3}}

Let now $0<\gamma<1$. By \eqref{sec:Mo.6} we represent the $\gamma$ power of 
the wealth process as 
$$
 (X^{y}_\zs{T})^{\gamma}=x^{\gamma}
\,e^{\gamma R_\zs{T}+\gamma F_\zs{T}(y)}\,\cE_\zs{T}(\gamma y)\,,
$$
where
\begin{equation}\label{sec:Pr.1}
 F_\zs{T}(y)=(\theta,y)_\zs{T}-\frac{1-\gamma}{2}\|y\|^2_\zs{T}\,.
\end{equation}

Moreover, we introduce the measure (generally non probability) by the following
Radon-Nikodym density
$$
\frac{\d\wt{\P}}{\d \P}=\cE_\zs{T}(\gamma y) 
\,.
$$
By denoting $\wt{\E}$ the expectation with respect to this measure we get
that
\begin{equation}\label{sec:Pr.2}
\E
(X^{y}_\zs{T})^{\gamma}
=x^{\gamma}\,
e^{\gamma R_\zs{T}}\,
\wt{\E}
e^{\gamma F_\zs{T}(y)}\,.
\end{equation}
Note that, if $\|\theta\|_\zs{T}=0$ then
$$
\E
(X^{y}_\zs{T})^{\gamma}
=x^{\gamma}\,
e^{\gamma R_\zs{T}}\,
\wt{\E}\,
e^{-\frac{\gamma(1-\gamma)}{2}\|y\|^2_\zs{T}}\,.
$$
Taking into account that for any process $y$ from $\cY$
$$
\E
\cE_\zs{T}(\gamma y) \le 1
$$
we get for any $y\in\cY$
$$
\E
(X^{y}_\zs{T})^{\gamma}
\le x^{\gamma}\,
e^{\gamma R_\zs{T}}
$$
with the equality if and only if $y_\zs{t}=0$.

Therefore, in the sequel we assume that $\|\theta\|_\zs{T}>0$.
Now we shall consider the almost sure optimisation problem for the function $F_\zs{T}(\cdot)$.
First, we consider this contrained the last time moment $t=T$, i.e.
\begin{equation}\label{sec:Pr.3}
\sup_\zs{y\in\cY}F_\zs{T}(y)
\quad\mbox{subject to}\quad
\frac{{\rm VaR}^*_\zs{T}(x,y,\alpha)}{\zeta_\zs{T}(x)}\,\le\,1
\quad\mbox{a.s..}
\end{equation}
This constraint is equivalent to 
$$
\frac{1}{2}\|y\|^2_\zs{T}
-\tau^{*}_\zs{T}\|y\|_\zs{T}
-(\theta,y)_\zs{T}
\le -\ln(1-\zeta)=:a_\zs{max}\,.
$$
By fixing the the quantile as $\tau^*_\zs{T}=-\beta$ for some 
$\beta\ge |z_\zs{\alpha}|$ and denoting
$$
K_\zs{T}(y)=\frac{1}{2}\|y\|^2_\zs{T}
+\beta\|y\|_\zs{T}
-(\theta,y)_\zs{T}
$$
we will consider more general problem than \eqref{sec:Pr.3}, i.e. we will find 
the optimal solution in the Hilbert space $\L_\zs{2}[0,T]$, i.e.
$$
\sup_\zs{y\in\L_\zs{2}[0,2]}F_\zs{T}(y)
\quad\mbox{subject to}\quad
K_\zs{T}(y)\le a_\zs{max}\,.
$$
To resolve this problem we have to resolve the following one
\begin{equation}\label{sec:Pr.4}
\sup_\zs{y\in\L_\zs{2}[0,T]} F_\zs{T}(y)
\quad\mbox{subject to}\quad
K_\zs{T}(y)=a
\end{equation}
for some parameter $0\le a\le a_\zs{max}$.
We use the Lagrange multiplicators method, i.e.
we pass to the Lagrange cost function $H_\zs{\lambda}(y)=F_\zs{T}(y)-\lambda K_\zs{T}(y)$
and we have to resolve the optimisation problem for this function, i.e.
\begin{equation}\label{sec:Pr.5}
\max_\zs{y\in\L_\zs{2}[0,T]}
H_\zs{\lambda}(y)\,.
\end{equation}
In this case
$$
H_\zs{\lambda}(y)
=
-\frac{\lambda+1-\gamma}{2}\|y\|^2_\zs{T}
+(1+\lambda)(\theta,y)_\zs{T}
-\lambda\beta\|y\|_\zs{T}
\,,
$$
where $\lambda$ is Lagrange multiplicator. It is clear that $\lambda>\gamma-1$. Since
the problem \eqref{sec:Pr.5} has no finite solution for $\lambda\le\gamma-1$, i.e.
$$
\max_\zs{y\in\L_\zs{2}[0,T]}
H_\zs{\lambda}(y)=+\infty\,.
$$
to this end we calculate the G\^ateau derivative, i.e.
$$
\D_\zs{\lambda}(y,h)=\lim_\zs{\delta\to 0}
\frac{H_\zs{\lambda}(y+\delta h)-H_\zs{\lambda}(y)}{\delta}\,.
$$
It is easy to check directly that for any function $y$ from $\L_\zs{2}[0,T]$
with $\|y\|_\zs{T}>0$ 
$$
\D_\zs{\lambda}(y,h)=
\int^T_\zs{0}h'_\zs{t}
\left(
(1+\lambda)\theta_\zs{t}
-(1-\gamma+\lambda)y_\zs{t}
-\lambda\beta
\ov{y}_\zs{t}
\right)
\d t
$$
with $\ov{y}_\zs{t}=y_\zs{t}/\|y\|_\zs{T}$.
Moreover, if $\|y\|_\zs{T}=0$, then
$$
\D_\zs{\lambda}(y,h)=
(1+\lambda)
\int^T_\zs{0}
h'_\zs{t}
\theta_\zs{t}
\d t
-\lambda\beta\|h\|_\zs{T}\,.
$$
It is clear that $D_\zs{\lambda}(y,h)\neq 0$ for 
$h_\zs{t}=-\sign(\lambda)\theta_\zs{t}$. Therefore, to resolve the equation
\begin{equation}\label{sec:Pr.6}
D_\zs{\lambda}(y,h)=0
\end{equation}
for all $h\in\L_\zs{2}[0,T]$ we assume that $\|y\|_\zs{T}>0$.
This implies
$$
(1+\lambda)\theta_\zs{t}
-(1-\gamma+\lambda)y_\zs{t}
-\lambda\beta
\ov{y}_\zs{t}=0\,,
$$
i.e.
$$
y_\zs{t}=
\frac{(1+\lambda)\|y\|_\zs{T}}{\lambda\beta+(1+\lambda-\gamma)\|y\|_\zs{T}}
\theta_\zs{t}\,.
$$
Therefore,
\begin{equation}\label{sec:Pr.6-1}
y^{\lambda}_\zs{t}=\frac{\psi(\lambda)}{\|\theta\|_\zs{T}}\,\theta_\zs{t}
\quad
\mbox{with}
\quad
\psi(\lambda)=
\frac{\|\theta\|_\zs{T}+\lambda(\|\theta\|_\zs{T}-\beta)}{1-\gamma+\lambda}\,.
\end{equation}
The coefficient $\psi$ must be positive, i.e. 
\begin{equation}\label{sec:Pr.7}
\gamma-1<
\lambda<
\frac{\|\theta\|_\zs{T}}{(\beta-\|\theta\|_\zs{T})_\zs{+}}\,.
\end{equation}

Now we have to verify that the solution of the equation \eqref{sec:Pr.6} gives
the maximum solution for the problem \eqref{sec:Pr.5}. To end this for any function
$y$ from $\L_\zs{2}[0,T]$ with $\|y\|_\zs{T}>0$ we set
$$
\Delta_\zs{\lambda}(y,h)=H_\zs{\lambda}(y+h)-H_\zs{\lambda}(y)-
D_\zs{\lambda}(y,h)\,.
$$
Moreover, by putting
\begin{equation}\label{sec:Pr.8}
\delta(y,h)=
\|y+h\|_\zs{T}
-
\|y\|_\zs{T}
-
(h,\ov{y})_\zs{T}
\,,
\end{equation}
we obtain 
$$
\Delta_\zs{\lambda}(y,h)
=
-\frac{\lambda+1-\gamma}{2}\|h\|^2_\zs{T}
-\lambda\beta
\delta(y,h)\,,
$$
Now Lemma~\ref{Le.sec:A.1} implies that the function $\Delta(y,h)\le 0$ for all 
$h\in\L_\zs{2}[0,T]$. Therefore the solution of the equation \eqref{sec:Pr.6}
gives the solution for the problem \eqref{sec:Pr.5}. 

Now we chose the lagrange multiplicator $\lambda$ to satisfy the condition in 
\eqref{sec:Pr.4}, i.e.
$$
K_\zs{T}(y^\lambda)=a\,,
$$
i.e.
$$
\psi^2(\lambda)+2\psi(\lambda)(\beta-\|\theta\|_\zs{T})=2a\,,
$$
i.e.
$$
\wt{\psi}(a)=
\psi(\lambda(a))=\sqrt{2a+(\beta-\|\theta\|_\zs{T})^2}
-(\beta-\|\theta\|_\zs{T})
$$
with
$$
\lambda=\lambda(a)=
\frac{\|\theta\|_\zs{T}+(1-\gamma)(\beta-\|\theta\|_\zs{T})}{\sqrt{2a+(\beta-\|\theta\|_\zs{T})^2}}
-1+\gamma\,.
$$
One can check directly that the function $\lambda(a)$ satisfies the condition \eqref{sec:Pr.7} for any $a>0$. This means
that the solution for the problem \eqref{sec:Pr.4} is given by the function 
$$
\wt{y}^a_\zs{t}=y^{\lambda(a)}_\zs{T}=\frac{\wt{\psi}(a)}{\|\theta\|_\zs{T}}\theta_\zs{t}\,.
$$
Now to chose the parameter $0<a\le a_\zs{max}$ in \eqref{sec:Pr.4} we 
have to maximize the function \eqref{sec:Pr.1}, i.e.
$$
\max_\zs{0\le a\le a_\zs{max}}
\,F_\zs{T}(\wt{y}^a)\,.
$$
Note that
$$
F_\zs{T}(\wt{y}^a)=
G(\wt{\psi}(a))
\quad\mbox{with}\quad
G(\psi)=
\psi\|\theta\|_\zs{T}
-(1-\gamma)
\frac{\psi^2}{2}\,.
$$
Moreover, note that for any $a>0$
and $\beta\ge |z_\zs{\alpha}|$
$$
\wt{\psi}(a)
\le g(a)\,,
$$
where the function $g$ is defined in \eqref{sec:Op.11}. Therefore,
$$
\max_\zs{0\le a\le a_\zs{max}}\,
F_\zs{T}(\wt{y}^a)\,
\le\,
\max_\zs{0\le a\le a_\zs{max}}\,
G(g(a))
= G(g(a^{*}))\,,
$$
where $a^{*}$ is defined in \eqref{sec:Op.14}. To obtain here the equality we take in
\eqref{sec:Pr.6-1} $\beta=|z_\zs{\alpha}|$. Thus, the function \eqref{sec:Op.15}
is the solution of the problem  \eqref{sec:Pr.3}. Now to pass to the problem
\eqref{sec:Op.10} we have to check the condition \eqref{sec:Op.9} for the function
 \eqref{sec:Op.15}. To this end note that
$$
\frac{1}{2}\|y^{*}\|^{2}_ \zs{t}
+|z_\zs{\alpha}| \|y^{*}\|_\zs{t}
-(\theta,y^{*})_\zs{t}=
\int^{t}_\zs{0}\,\omega(s)\,\d s\,,
$$
where
$$
\omega_\zs{s}=|\theta_\zs{s}|^{2}
\left(
\frac{(g^{*})^{2}}{2\|\theta\|^{2}_\zs{T}}
+
\frac{g^{*}(|z_\zs{\alpha}|-2\|\theta\|_\zs{s})}{2\|\theta\|_\zs{T}\|\theta\|_\zs{s}}
\right)\,.
$$
taking into account here the condition 
$|z_\zs{\alpha}|\ge\|\theta\|_\zs{T}$ we obtain $\omega_\zs{t}\ge 0$, i.e.
\begin{align*}
&\frac{1}{2}\|y^{*}\|^{2}_ \zs{t}
+|z_\zs{\alpha}| \|y^{*}\|_\zs{t}
-(\theta,y^{*})_\zs{t}\\[2mm]
&\,\le\,
 \frac{1}{2}\|y^{*}\|^{2}_ \zs{T}
+
|z_\zs{\alpha}| \|y^{*}\|_\zs{T}
-(\theta,y^{*})_\zs{T}\\[2mm]
&=
a^{*}
\le -\ln (1-\zeta)\,.
\end{align*}
This implies immediately that the function \eqref{sec:Op.15}  
is a solution of the problem \eqref{sec:Op.10}.

\halmos

\subsection{Proof of Theorem~\ref{Th.sec:Op.4}}

Let now $\gamma=1$. Note that in this case we can obtain the following  upper bound:
$$
\E\, X^{y}_\zs{T}\le x e^{R_\zs{T}}\,
\E\,e^{\|\theta\|_\zs{T} \|y\|_\zs{T}}\,\cE_\zs{T}(y)\,.
$$ 
Obviously, that if $\|\theta\|_\zs{T}=0$ than we obtain here equality if and only if $y=0$.
Let now $\|\theta\|_\zs{T}>0$. Note that the condition
\begin{equation}\label{sec:Pr.9}
K_\zs{T}(y)\le a_\zs{max}
\end{equation}
implies $\|y\|_\zs{T}\le g(a_\zs{max})$. Thus, for any function $(y_\zs{t})_\zs{0\le t\le T}$
satisfying the condition we have
$$
\E\, X^{y}_\zs{T}\le x e^{R_\zs{T}+g(a_\zs{max})\|\theta\|_\zs{T}}\,.
$$
Moreover, the function \eqref{sec:Op.18} transforms this inequality in the equality. 
By the same way as in the proof of Theorem~\ref{Th.sec:Op.4} we check that the function 
\eqref{sec:Op.18} satisfies the condition \eqref{sec:Op.9}.

\halmos

\newpage
\renewcommand{\theequation}{A.\arabic{equation}}
\renewcommand{\thetheorem}{A.\arabic{theorem}}
\renewcommand{\thesubsection}{A.\arabic{subsection}}
\section{Appendix}\label{sec:A}
\subsection{Properties of the function \eqref{sec:Pr.8}}

\begin{lemma}\label{Le.sec:A.1}
Assume that $y\in \L_\zs{2}[0,T]$ with $\|y\|_\zs{T}>0$.
Then for every $h\in\L_\zs{2}[0,T]$ the function \eqref{sec:Pr.8} is positive, i.e. 
$\delta(y,h)\ge 0$.
\end{lemma}

\noindent{\bf Proof. \,}
Obviously, if $h\equiv ay$ for some $a\in\bbr$, then $\delta(y,h)=(|1+a|-1-a)\|y\|_\zs{T}\ge 0$.
Let now the functions $h$ and $y$ be linearly independent. Then
$$
\delta(y,h)=\frac{2(y\,,\,h)_\zs{T}+\|h\|^2_\zs{T}}{\|y+h\|_\zs{T}+\|y\|_\zs{T}}
-(\ov{y},h)_\zs{T}
=\frac{\|h\|^2_\zs{T}-(\ov{y},h)_\zs{T}((\ov{y},h)_\zs{T}+\delta(y,h) )}
{\|y+h\|_\zs{T}+\|y\|_\zs{T}}\,.
$$
It is clear that for all $h$
$$
\|y+h\|_\zs{T}+\|y\|_\zs{T}+(\ov{y},h)_\zs{T}\ge 0
$$
with equality if and only if $h\equiv ay$ for some $a\le -1$.\\
Therefore, if the functions $h$ and $y$ are linearly independent, then
$$
\delta(y,h)=
\frac{\|h\|^2_\zs{T}-(\ov{y}\,,\,h)^2_\zs{T}}{\|y+h\|_\zs{T}+\|y\|_\zs{T}+(\ov{y},h)_\zs{T}}\,
\ge\,0\,.
$$

\halmos

\end{document}